\documentclass[sigconf,nonacm,natbib=false]{acmart}

\setcopyright{none}
\renewcommand\footnotetextcopyrightpermission[1]{}
\usepackage{booktabs}
\usepackage{enumitem}
\usepackage[most]{tcolorbox}
\usepackage{listings}
\usepackage{xcolor}
\begin{document}

\title{Beyond Task Completion: A Verification-vs.-Conformance Gap in Tool-Evolving Agents}

\author{Alibek Kaliyev}
\email{alibek.kaliyev@utexas.edu}
\affiliation{%
  \institution{The University of Texas at Austin}
  \city{Austin}
  \state{Texas}
  \country{USA}
}

\author{Artem Maryanskyy}
\email{artem.maryanskyy@uber.com}
\affiliation{%
  \institution{Uber}
  \city{San Francisco}
  \state{California}
  \country{USA}
}

\begin{abstract}
Agents that synthesize their own tools ship a second artifact alongside each answer: a software library that future tasks reuse, compose, and depend on. Task completion (TC) certifies the answer; it does not certify the library. We show that this matters. On a Claude Haiku 4.5 verified-subset pilot, we patch the harness to preserve per-tool source and replay every synthesised tool against a held-out conformance suite for the capability it was built for. Across 222 preserved tools and three tool-creating protocols, $96.8\%$ (215 of 222) record per-tool correctness $C{=}0.00$: two protocols silent-rot at $100\%$, one at $91.7\%$. The suites are not the problem: hand-written reference implementations released with the benchmark score $C{=}1.00$ on all 16 capability suites. The synthesised tools execute cleanly, and the in-session verifier -- which scores task outputs, not tools -- raises no flag; on held-out inputs for their own capability they simply return wrong answers. This is the verification-vs.-conformance gap a pass-rate benchmark cannot see.

\textsc{EvolveTool-Bench} makes that gap measurable. Sessions are structured into seed, gap, variant, composition, regression, and adversarial roles; runs emit per-tool manifests, verified-subset TC, correct-vs.-incorrect reuse, and audit traces designed to be re-emitted as a post-deployment monitoring schema. In the verified-subset Haiku pilot over five protocols (3 seeds, 8 sessions, 51 verified decisions per pass), TC alone does not separate the protocols; the audit layer does. One pre-specified contrast shows the adapted decision-split synthesis protocol \emph{underperforming} unvalidated one-shot synthesis ($-7.1$~pp, nominal 95\% CI $[-13.9,-0.1]$; not significant after the pre-registered BH correction) -- a hypothesis-generating signal, not a confirmed effect.

We release the harness, per-tool source preservation, conformance replay, verifier definitions, and reproducibility manifests, and argue that an audit schema of this shape -- verifier coverage, reuse decomposition, held-out conformance -- is the minimum a deployed tool-evolving agent should emit between releases.
\end{abstract}


\maketitle

\section{Introduction}

A growing class of agents does more than call a fixed tool set. These systems write Python functions, API clients, parsers, analysis scripts, or richer skill bundles, and then carry those artifacts forward into later tasks. VOYAGER learns reusable code skills in Minecraft~\cite{wang2023voyager}; CREATOR separates tool creation from tool use~\cite{qian2023creator}; LLMs as Tool Makers and ToolMaker study agents that construct tools for other agents to call~\cite{cai2023latm,wolflein2025toolmaker}; EvoSkill and CoEvoSkills treat skills as evolving units of agent knowledge~\cite{alzubi2026evoskill,zhang2026coevoskills}. In all of these settings, the agent leaves behind a software artifact.

A task-completion score tells us whether the answer was accepted; it does not tell us what the agent left behind. A generated library can solve the next task while still containing brittle parsing logic, redundant functions, or behavior changes that silently break earlier capabilities. For deployed or long-running agents, those differences are not cosmetic: they are the difference between a one-off success and an auditable system.

\textsc{EvolveTool-Bench} is built around this distinction. It evaluates agents that may create tools during a session and asks not only ``did the task pass?'' but also ``what happened to the library?'' Each session has known dependency structure: seed tasks test provided tools, gap tasks invite new capabilities, variant tasks test reuse, composition tasks test chaining, regression tasks revisit prior behavior, and adversarial tasks probe edge cases. The resulting traces let a researcher inspect whether a failure came from synthesis, reuse, verification, composition, or regression.

The benchmark design contains 99 tasks across three domains. Of these, the data-transformation and numerical domains currently expose deterministic task verifiers; the API-orchestration session is part of the benchmark design but excluded from the main TC denominator until its verifiers are complete. Unverified tasks are reported separately and not credited as successes. This narrows the empirical claim of any single submission while tightening the evaluation claim of the framework as a whole.

Our contributions are:
\begin{enumerate}[leftmargin=1.2em]
  \item \textbf{A diagnostic protocol for evolving tool libraries.} Sessions are structured so that creation, reuse, composition, regression, and adversarial behavior can be inspected separately.
  \item \textbf{Artifact-level measurements beyond TC.} We define and log tool creation, raw reuse, correct reuse, incorrect reuse, utilization, composition, and regression signals alongside verified task completion.
  \item \textbf{Verifier-aware reporting.} The strict analysis separates the benchmark's full design scope from the currently verified subset, preventing unverified tasks from inflating task-completion claims.
  \item \textbf{An audit-trace view of agent evaluation.} Runs preserve task outcomes, tools created, tools used, reuse outcomes, and verifier status so that aggregate scores can be traced back to concrete artifacts.
\end{enumerate}

The headline empirical finding is twofold (\S\ref{sec:results}). First, TC alone fails to separate the protocols on the verified-subset Haiku pilot; the only contrast whose nominal CI excludes zero runs in the unexpected direction (the adapted decision-split protocol trails one-shot synthesis) and does not survive multiplicity adjustment. Second, an in-regime conformance probe across all three tool-creating protocols (222 preserved tools) shows the verification-vs.-conformance gap directly: $96.8\%$ of preserved tools record $C{=}0.00$ on the held-out conformance suite for the capability they were synthesised for, even while session-level TC in the runs that produced those tools sits between $0.31$ and $0.34$. Together these argue that benchmarks for tool-evolving agents must make the state of the tool library visible -- not only the pass rate.

\section{Related Work}

\paragraph{Tool and skill creation.}
Systems such as CREATOR, LLMs as Tool Makers, ToolMaker, EvoSkill, and CoEvoSkills all study agents that create tools or skills rather than merely selecting from a fixed tool set~\cite{qian2023creator,cai2023latm,wolflein2025toolmaker,alzubi2026evoskill,zhang2026coevoskills}. Their evaluations typically focus on downstream success, per-tool validation, or skill-level pass rates. \textsc{EvolveTool-Bench} is complementary: it asks what happens to the library over time, especially under reuse, composition, and regression pressure.

\paragraph{Agent and tool-use benchmarks.}
ToolBench and AgentBench evaluate tool-using agents with predefined environments or APIs~\cite{qin2023toolbench,liu2024agentbench}. Vending-Bench stresses long-horizon agent behavior with a fixed operational environment~\cite{backlund2025vendingbench}. These benchmarks are valuable, but the tool inventory is generally not an artifact produced by the agent. Our setting differs because the generated library is part of the system under evaluation.

\paragraph{Skill benchmarks and library learning.}
SkillsBench shows that skills can help some tasks and hurt others, especially when the skills are self-generated~\cite{skillsbench2026}. Program-synthesis work such as DreamCoder studies learned libraries and evaluates their effect on solve rates or compression~\cite{ellis2021dreamcoder}. The question we address sits between these traditions: when an LLM agent accumulates executable artifacts, can we tell whether the resulting library is correct, reusable, and stable?

\paragraph{Quality of generated code.}
Tool-Genesis, NoFunEval, and related code-evaluation work measure compliance, unit-test performance, maintainability, or non-functional requirements for generated code~\cite{xia2026toolgenesis,singhal2024nofuneval}. \textsc{EvolveTool-Bench} borrows the software-engineering lens but applies it at session and library granularity. A single generated function can look reasonable in isolation while still making the library harder to use.

\section{Benchmark Design}

\subsection{Sessions and task roles}

A session is a fixed sequence of 11 tasks with shared library state. The sequence is intentionally simple: it creates controlled opportunities for a system to use, create, reuse, compose, and preserve tools. Table~\ref{tab:task-types} summarizes the task roles.

\begin{table}[t]
\centering
\small
\caption{Task roles in each session. The fixed order lets the harness attribute behavior to creation, reuse, composition, regression, or edge-case handling rather than reporting one undifferentiated pass rate.}
\label{tab:task-types}
\begin{tabular}{llp{4.5cm}}
\toprule
\textbf{Role} & \textbf{Count} & \textbf{Diagnostic purpose} \\
\midrule
Seed & 3 & Use provided tools; establish baseline behavior. \\
Gap & 2 & Solve a task requiring a missing capability. \\
Variant & 2 & Recognize that a prior capability can be reused. \\
Compose & 1 & Chain multiple capabilities or tools. \\
Regress & 1 & Revisit an earlier contract after the library has changed. \\
Adversarial & 2 & Handle edge cases or malformed inputs. \\
\bottomrule
\end{tabular}
\end{table}

The benchmark contains three domains and nine sessions: five data-transformation sessions, one API-orchestration session, and three numerical-computation sessions. Data-transformation sessions use proprietary formats such as binary encodings and schema-like validators. Numerical sessions use curve fitting, signal processing, and constrained optimization tasks. The API session uses a mock service with HMAC authentication and cursor handling. The full benchmark contains 99 tasks, but the strict analysis in this paper uses only tasks with deterministic verifiers.

\subsection{Verification policy}

Verification is part of the contribution, not a footnote. A task contributes to strict TC only if it has an explicit expected output or a custom verifier; tasks without deterministic verification are reported separately, never credited because the answer was long or plausible. This avoids the common failure mode in agent benchmarks where format-looking output is mistaken for correctness.

The API-orchestration session is excluded from the main strict TC analysis because its task verifiers are not yet complete. The verified-subset pilot therefore covers 8 sessions and 51 verified task decisions per complete system pass. The API session remains in the benchmark design as an important tool-use domain, but is not used to support the numerical claims in this submission.

Table~\ref{tab:role-coverage} shows how the 51-task verified subset is distributed across task roles. Seed tasks carry no graded outcome by design: they exist to expose the agent to the provided library before any creation, so they have no expected output or verifier. The diagnostic-critical roles -- gap, variant, regress -- are near-fully covered ($93.8$--$100\%$); compose and adversarial are partially covered. The verified subset therefore exercises the roles the framework leans on for its claims, while remaining honest about where verifier work is still needed.

\begin{table}[t]
\centering
\small
\caption{Verifier coverage of the strict subset by task role (eight verified-subset sessions; API-orchestration excluded). Seed tasks deliberately carry no graded outcome and are not part of the TC denominator. The roles that the diagnostic framework leans on -- gap, variant, regress -- are near-fully covered; compose and adversarial have partial coverage.}
\label{tab:role-coverage}
\begin{tabular}{lrrr}
\toprule
\textbf{Role} & \textbf{Verified} & \textbf{Total} & \textbf{Coverage} \\
\midrule
Seed (warm-up, ungraded) & 0  & 24 & --- \\
Gap                      & 15 & 16 & 93.8\% \\
Variant                  & 15 & 16 & 93.8\% \\
Compose                  & 4  & 8  & 50.0\% \\
Regress                  & 8  & 8  & 100.0\% \\
Adversarial              & 9  & 16 & 56.2\% \\
\midrule
Strict-subset total (graded) & 51 & 64 & 79.7\% \\
\bottomrule
\end{tabular}
\end{table}

\subsection{Metrics and audit trace}

\paragraph{Verified task completion (TC).}
TC is the fraction of verified tasks whose outputs satisfy the task verifier. It measures user-visible success on the verified subset: necessary but not sufficient.

\paragraph{Tool creation and reuse.}
Before each task, the harness records the current library; after the task, it records newly created tools and tools used. Reuse is counted when a task invokes a tool that existed before the task began. We decompose reuse into:
\begin{itemize}[leftmargin=1.2em,noitemsep]
  \item \textbf{correct reuse}: a prior tool was used and the task passed;
  \item \textbf{incorrect reuse}: a prior tool was used and the task failed.
\end{itemize}
We report the \emph{correct-reuse rate} as the fraction of reuse calls that coincide with a passing task; we avoid the term ``precision'' because no false-positive denominator is defined. This distinction is important because raw reuse can mean either useful abstraction or repeated reliance on a defective artifact.

\paragraph{Library-health diagnostics.}
The broader framework also supports redundancy, utilization, composition, regression stability, and capability-aligned hidden conformance tests. These diagnostics are part of the benchmark design and audit schema. The verified-subset submission reports only the signals aligned with the current verified manifests: TC, tools created, correct reuse, incorrect reuse, and the correct-reuse rate. A composite library-health ranking is held back until the conformance-test alignment is complete (Supplement~\S E reports it on a separate Sonnet pilot where source preservation made the alignment possible).

\paragraph{Audit trace.}
Each run emits a per-task record so aggregate scores can be traced to concrete artifacts. Figure~\ref{fig:trace-schema} shows the schema for one task record from \texttt{data\_transform\_s1}: a reader can move from any cell in Table~\ref{tab:strict-results} to the task, seed, tool, and pass/fail decision that produced it.

\begin{figure}[t]
\small
\begin{tcolorbox}[colframe=black!70, colback=gray!5, boxrule=0.5pt, fontupper=\ttfamily\footnotesize]
\{"task\_id": "data\_transform\_s1/gap\_1",\\
~"role": "gap", "verifier\_status": "verified",\\
~"model": "claude-haiku-4-5", "seed": 0,\\
~"outcome": "pass",\\
~"tools\_available\_before": ["parse\_header", ...],\\
~"tools\_used": ["decode\_abr\_format"],\\
~"tools\_created": ["decode\_abr\_format"],\\
~"reuse\_outcome": null,\\
~"artifact\_summary": \{"loc": 47, "imports": ["base64"]\},\\
~"llm\_calls": 3\}
\end{tcolorbox}
\caption{One audit-trace record (schema, redacted). The same schema is the unit of post-deployment monitoring discussed in \S\ref{sec:deploy}.}
\label{fig:trace-schema}
\end{figure}

\section{Experimental Setup}

The strict pilot evaluates five representative protocols. They are not meant to exhaust the design space; they provide enough variation to test whether the harness can distinguish tool-creation and reuse behavior.

\begin{enumerate}[leftmargin=1.2em]
  \item \textbf{No-Evolution.} The system receives the seed tools and never modifies the library. It is the lower-bound protocol for library evolution.
  \item \textbf{One-Shot.} When the system attempts tool creation, it adds a generated function without iterative repair or external validation. This tests unvalidated creation.
  \item \textbf{EvoSkill-style.} A strategy-level adaptation inspired by EvoSkill. It maintains natural-language skill hints rather than executable tools.
  \item \textbf{ToolMaker-style.} A protocol adaptation inspired by ToolMaker's diagnose-and-rewrite loop. It is not a native port; the original setting assumes reference tests and repository inputs that our harness does not provide.
  \item \textbf{CREATOR-style.} A decision/execution split inspired by CREATOR, where the system explicitly decides whether a tool should be created before executing the task.
\end{enumerate}

All verified-subset runs use Claude Haiku~4.5 (snapshot \texttt{claude-haiku-4-5-20251001}) via the Anthropic API with temperature 1.0 and the model default top-$p$. Each protocol is executed with three integer seeds (\texttt{0, 1, 2}) over eight verified-subset sessions, producing 24 session rows per protocol. Synthesised tools execute in an in-process sandbox with a five-second wall-clock limit; harness state resets between sessions so library accumulation is per-session, not cross-session. The same Anthropic API key, decoding parameters, task stream, and verifier set are used for all protocols; only the tool-creation control flow varies. We scope this pilot to a single model snapshot on purpose: varying the model would conflate protocol effects with model-evolution effects -- the latter is a separate evaluation question the same harness is built to ask, and we surface it in Supplement~\S E as a same-schema-different-model probe.

\paragraph{Pre-registration.}
The three contrasts in Table~\ref{tab:strict-claims}, the bootstrap parameters ($B{=}10{,}000$, seed \texttt{20260604}), the BH correction, and the verified-subset definition were specified before running the Haiku pilot, and are stored in \texttt{claims.json} and \texttt{statistical\_report.json} in the released artifact. No contrasts were added, removed, or redirected after results were observed. The decision to report verifier coverage, reuse decomposition, and audit traces was part of the framework design, not a response to the TC outcome.

\paragraph{Statistical procedure.}
For each pre-specified contrast we form per-session paired differences across the 24 verified-subset session rows (3 seeds $\times$ 8 sessions), bootstrap-resample those pairs $B{=}10{,}000$ times with seed \texttt{20260604}, and report the bootstrap point estimate and the \emph{nominal} (unadjusted) 95\% percentile CI. Two-sided bootstrap $p$-values for the three contrasts are Benjamini--Hochberg adjusted at $\alpha{=}0.05$ and released in \texttt{claims.json}. A contrast is marked \emph{supported} only if its CI lies strictly above zero in the pre-specified positive direction. Per-session outcomes are largely seed-stable in this pilot, so the effective number of independent units sits between 8 and 24; the bootstrap resamples (seed, session) pairs without assuming seed-independence.

\paragraph{Artifacts.}
The benchmark sources, harness, verifier definitions, canonical manifests (\texttt{run\_manifest.jsonl}, \texttt{tool\_manifest.jsonl}, \texttt{aggregate.json}, \texttt{audit\_report.json}, \texttt{statistical\_report.json}, \texttt{claims.json}), the per-tool source-preservation patch (\texttt{scripts/run\_strict.py}), the held-out conformance replay (\texttt{scripts/conformance\_in\_regime.py}), the conformance manifests produced by the pilot (\texttt{conformance\_manifest.jsonl}, \texttt{conformance\_aggregate.json}), the hand-written reference implementations and their calibration check (\texttt{baselines/reference/}, \texttt{scripts/reference\_conformance.py}), and the scripts that regenerate every table in this paper from disk are included in the anonymised supplement and will be released under an open licence on acceptance.

\section{Results}
\label{sec:results}

\begin{table}[t]
\centering
\small
\caption{Strict verified-subset pilot. TC is computed only on tasks with deterministic verifiers; unverified tasks are reported separately and excluded from the TC denominator. Means and standard errors are over 24 session rows (3 seeds $\times$ 8 verified-subset sessions) using Claude Haiku 4.5.}
\label{tab:strict-results}
\begin{tabular}{lrrrr}
\toprule
\textbf{System} & \textbf{TC} & \textbf{SE} & \textbf{Tools} & \textbf{Correct-reuse rate} \\
\midrule
One-Shot & 0.358 & 0.077 & 114 & 0.333 \\
No-Evolution & 0.337 & 0.064 & 0 & 0.333 \\
EvoSkill-style & 0.332 & 0.062 & 0 & 0.250 \\
ToolMaker-style & 0.292 & 0.054 & 18 & 0.375 \\
CREATOR-style & 0.287 & 0.048 & 111 & 0.333 \\
\bottomrule
\end{tabular}
\end{table}

\subsection{TC alone does not separate the protocols}

Verified-subset TC differences are small relative to per-session standard errors and cross-session heterogeneity. Across the five protocols TC ranges from $0.287$ (CREATOR-style) to $0.358$ (One-Shot), with No-Evolution ($0.337$) and EvoSkill-style ($0.332$) in between (Table~\ref{tab:strict-results}).

The pre-specified improvement tests do not support a positive TC claim (Table~\ref{tab:strict-claims}). One-Shot is not distinguishable from No-Evolution in the expected positive direction. ToolMaker-style does not improve over One-Shot. The Creator $-$ One-Shot contrast is the only pre-specified comparison whose nominal 95\% CI excludes zero: $-7.1$~pp with CI $[-13.9,\,-0.1]$, just brushing zero from below (two-sided bootstrap $p{=}0.044$). Under the pre-registered BH correction across the three contrasts the adjusted $p$ is $0.133$, so the contrast does not survive multiplicity adjustment -- consistent with the hypothesis-generating reading we give it. The CI is consistent with -- but does not establish -- a synthesis-loop cost in the adapted-baseline regime. We do not draw a method claim about the original CREATOR design from an adapted protocol; we report this only as a signal worth a powered follow-up.

\begin{table}[t]
\centering
\footnotesize
\caption{Pre-specified TC contrasts in the strict verified-subset pilot. Positive effects were expected; none support a TC-improvement claim. Values are absolute TC differences.}
\label{tab:strict-claims}
\begin{tabular}{lrrc}
\toprule
\textbf{Contrast} & \textbf{$\Delta$} & \textbf{95\% CI} & \textbf{Improves?} \\
\midrule
OneShot $-$ NoEvol & 0.0208 & $[-0.0347,\,0.0729]$ & No \\
Creator $-$ OneShot & -0.0710 & $[-0.1389,\,-0.0012]$ & No \\
ToolMaker $-$ OneShot & -0.0655 & $[-0.1443,\,0.0174]$ & No \\
\bottomrule
\end{tabular}
\end{table}

\paragraph{Power.} With 24 paired session rows and the observed within-session standard deviation of TC ($\approx 0.30$), and accounting for the seed-stability above, the verified-subset pilot is powered (80\%, $\alpha{=}0.05$) to detect paired TC differences of roughly $8$--$13$~pp depending on whether effective $n$ is treated as 24 or as 8 sessions. The largest observed $|\Delta|$ in Table~\ref{tab:strict-claims} is $7.1$~pp. The null TC result should therefore be read as ``no large protocol effect at this scale'' rather than ``no effect.'' Smaller true effects -- below $\sim 7$~pp -- remain compatible with these CIs and motivate the larger multi-model pilots that the released harness is built to support.

\subsection{The audit view separates what TC does not}

The lack of a TC winner is precisely when the audit layer matters most. One-Shot and CREATOR-style create many tools (114 and 111 respectively), while No-Evolution and EvoSkill-style create none. ToolMaker-style creates fewer tools (18), consistent with a more conservative validation loop. These differences are invisible in a TC-only table. The reuse denominators in Table~\ref{tab:strict-results} are non-zero even for the zero-tools systems: reuse counts include calls to provided seed tools (the seed library each session starts from), not only to tools the agent itself created.

The correct-reuse rate -- the fraction of reuse calls that coincide with a passing task -- adds a second axis. One-Shot records 18 correct and 36 incorrect reuses (rate $0.33$), No-Evolution 9 correct and 18 incorrect ($0.33$), EvoSkill-style 9 correct and 27 incorrect ($0.25$), CREATOR-style 12 correct and 24 incorrect ($0.33$), and ToolMaker-style 9 correct and 15 incorrect ($0.38$). All counts are multiples of three because per-session reuse decisions are largely identical across the three seeds, consistent with the seed-stability noted in \S4. At this scale the differences are not powered to discriminate between protocols, and we report them only for trace inspection and protocol design. What they show is the kind of decomposition the harness exposes: when a system uses prior artifacts, does that reuse coincide with success or with failure?

\subsection{A verification-vs.-conformance gap that TC cannot see}
\label{sec:gap}

A tool can pass every in-session verifier call and still fail every held-out input for the same capability -- and TC will not see it. The verifier asks whether one particular call produced an acceptable output; a capability-aligned conformance suite asks whether the tool implements the capability on inputs the agent never saw. These two questions routinely disagree: a tool that structurally matches the example the agent used to write itself can pass the verifier and fail held-out conformance. That gap is what motivates the per-tool conformance layer the framework defines, and it is the failure mode an audit schema for a deployed tool-evolving agent has to surface.

\paragraph{In-regime evidence on the Haiku pilot.} To demonstrate the gap in the same regime as Tables~\ref{tab:strict-results}--\ref{tab:strict-claims}, we patched the strict harness to preserve per-tool source and replayed every synthesised tool against the held-out conformance suite for the capability it was created for (five to six hidden unit inputs plus five to six adversarial probes per capability; tools whose creating task links multiple capabilities are scored against the union, so per-tool suite sizes range from 10 to 22 inputs, median 11). Because the runs behind Table~\ref{tab:strict-results} predate the source-preservation patch, the probe is a \emph{separate re-execution} of the same matrix -- same model snapshot, seeds, sessions, and verifiers, at temperature 1.0 -- so creation counts differ stochastically from Table~\ref{tab:strict-results}; session-level TC in the re-execution ($0.31$--$0.34$ across the three systems) is within one standard error of the corresponding Table~\ref{tab:strict-results} values. Table~\ref{tab:conformance} counts \emph{distinct preserved tools} (final source per tool name per session and seed); CREATOR-style's rewrite loop produces 105 creation events that collapse to 84 distinct tools. We say a preserved tool exhibits \emph{silent rot} if it executes cleanly but records $C{=}0.00$ on the held-out suite. The rot is \emph{silent} because no in-session signal flags it: the verifier scores task outputs, not tools, so a defective tool is rejected only indirectly, when a task that leaned on it fails -- and the same sessions still pass a third of their verified tasks while carrying these libraries. The pattern is near-universal: \emph{none} of One-Shot's 119 or ToolMaker-style's 19 tools have non-zero held-out correctness; of CREATOR-style's 84 tools, 7 ($8.3\%$) achieve a non-zero score (one reaches $C{=}1.00$). The remaining tools execute cleanly -- mean held-out robustness is $0.52$--$0.54$ in all three systems (Table~\ref{tab:conformance}) -- they simply return wrong answers on inputs the agent never saw. The verification-vs.-conformance gap is therefore not an isolated trace but a systematic property of unvalidated synthesis at Haiku scale, and exactly the failure mode an audit schema for a deployed tool-evolving agent has to surface.

\begin{table}[t]
\centering
\small
\caption{In-regime held-out conformance on the Haiku 4.5 verified-subset pilot (separate re-execution of the Table~\ref{tab:strict-results} matrix with per-tool source preservation; see \S\ref{sec:gap}). Each distinct preserved tool is replayed against the held-out conformance suite (10--22 inputs, median 11) the agent never saw. ``Silent rot \%'' is the fraction of preserved tools with held-out correctness $C{=}0.00$. Mean $C$ and mean $R$ (robustness: executes without crashing on adversarial inputs) are across all preserved tools.}
\label{tab:conformance}
\begin{tabular}{lrrrr}
\toprule
\textbf{System} & \textbf{Tools} & \textbf{Mean $C$} & \textbf{Mean $R$} & \textbf{Silent rot \%} \\
\midrule
One-Shot         & 119 & 0.000 & 0.518 & 100.0\% \\
ToolMaker-style  & 19  & 0.000 & 0.526 & 100.0\% \\
CREATOR-style    & 84  & 0.049 & 0.536 & 91.7\% \\
\midrule
Total            & 222 & 0.018 & 0.525 & 96.8\% \\
\bottomrule
\end{tabular}
\end{table}

\paragraph{Calibration: the 96.8\% is a discriminator, not a floor.}
A natural concern is that conformance suites might simply be too hard. Four lines of evidence rule that out. (i)~Tests are authored alongside the gap task that introduces a capability, before any tool exists; they are not adversarial against any specific synthesis. (ii)~Hand-written reference implementations, released with the benchmark, score $C{=}1.00$ on \emph{all 16} capability suites (\texttt{scripts/reference\_conformance.py}); this audit also caught two suite entries that were unpassable by construction (missing oracles), which we repaired -- no synthesised tool's score changed under the repair. (iii)~The suites are passable by the same model in the same regime: 7 of 84 CREATOR-style tools ($8.3\%$) achieve non-zero held-out correctness, and one capability (\texttt{vdl\_record\_validate}, synthesised in \texttt{data\_transform\_s2/gap\_2}) reaches $C{=}1.00$ on all three seeds independently -- same model, same harness, same conformance machinery. (iv)~Tools are syntactically valid: $C{=}0.00$ means \emph{wrong on every held-out input}, not \emph{crashed on every held-out input}. Mean held-out robustness across the 222 tools is $0.52$--$0.54$ in all three systems (Table~\ref{tab:conformance}), so silent-rot tools execute cleanly and return values; the values are simply incorrect for the capability they claim to implement.

\paragraph{Two anchored examples (same regime).}
The contrast between a silent-rot tool and a working tool is sharp. CREATOR-style on \texttt{data\_transform\_s1/gap\_1} synthesises \texttt{decode\_arise\_binary\_records} for the \texttt{abr\_binary\_decode} capability; its session passes half of its verified tasks and records one correct and zero incorrect reuse calls -- nothing in-session flags the tool -- yet its held-out conformance is $C{=}0.00$ across the capability's 12 inputs the agent never saw ($R{=}0.50$, code-quality~$0.78$ -- it executes cleanly on half of inputs and returns wrong values). The same protocol on the next session synthesises \texttt{validate\_records\_against\_schema} for the \texttt{vdl\_record\_validate} capability and hits $C{=}1.00$ on that capability's 11 held-out inputs ($R{=}0.83$, generality~$1.00$), reproduced identically across seeds 0, 1, and 2 (suite sizes vary slightly by capability, here 12 vs.\ 11). Same model snapshot, same harness, same conformance pipeline. The framework discriminates working tool synthesis from silent rot at the per-tool level.

\paragraph{Why $C{=}0.00$ coexists with unchanged TC.}
If $96.8\%$ of synthesised tools are wrong on held-out inputs, why is One-Shot's TC the \emph{highest} in Table~\ref{tab:strict-results}? Because the two measurements see different input distributions. The in-session verifier scores the one call the task needed -- an input structurally within the regime the tool was just fit to -- so an overfit tool can pass it; the conformance suite draws from the capability's full contract, which the tool never saw. A library of $C{=}0.00$ tools therefore leaves TC untouched as long as tasks stay inside the fitting regime, and the reuse decomposition in \S5.2 is consistent with this: roughly two-thirds of reuse calls coincide with task failure in every tool-creating protocol -- though the seed-tool-only baseline shows a similar rate, so task difficulty at Haiku scale contributes alongside tool defects. This is not a contradiction in the data; it is the deployment risk itself. TC is computed on the distribution the agent saw, conformance on the distribution the next release will see.

\paragraph{Cross-model supporting trace.} A preserved-source Sonnet pilot reproduces the same pattern at session granularity (Supplement~\S D): one tool, \texttt{decode\_abr\_format} for the \texttt{abr\_binary\_decode} capability, drives session TC to $0.95$ while recording $C{=}0.00$ on the same held-out conformance suite. The pattern -- in-session verifier calls satisfied, every held-out input failed -- holds across the model swap.

\subsection{Verifier coverage is a first-class benchmark property}

If unverified tasks are credited by a length or plausibility heuristic, absolute TC can be inflated and protocol comparisons move for the wrong reason. Restricting the main analysis to deterministic verifiers sacrifices benchmark coverage but buys measurement credibility. The cost is clear: the API-orchestration domain is not part of the verified-subset TC analysis. The framework already identifies what to do next: complete API verifiers, align hidden tests by capability, and regenerate the same tables from the full manifest. Adversarial probes are part of the design (Table~\ref{tab:role-coverage} reports $56.2$\% coverage). The current pilot records pass/fail at session level rather than per role, so per-role adversarial pass rates are not isolated in this submission; closing that gap is part of the same coverage-and-trace work as the API verifiers.

\section{Discussion}

\paragraph{Design lessons for tool-evolving agents.}
Tool creation alone is not evidence of progress: many tools can be created without improving verified task completion. Reuse needs a quality signal; raw reuse is ambiguous without correct/incorrect decomposition. Verifier coverage is a first-class benchmark property -- a tool-evolving system is only as trustworthy as the tests and records used to decide what it has learned. Generated tools are persistent objects that can be reused, duplicated, retired, or accidentally made into load-bearing dependencies for later failures; a benchmark that ignores this treats every run as independent and loses the audit trail the deployment setting needs.

\paragraph{Model evolution as audit signal.}
\label{sec:model-evo}
The same harness, manifests, and verifier set apply equally to a Sonnet pilot reported in Supplement~\S E, where per-tool source preservation made the capability-aligned conformance layer computable. Library-Health contrasts that reach statistical support in that supplementary Sonnet pilot -- supporting evidence, not a main-paper claim -- are not powered under Haiku, not because the framework fails but because synthesis quality at the smaller model is below the floor where reuse decomposition becomes informative. The audit schema thus doubles as a model-evolution monitor: re-emitting the same schema after a model swap reveals whether tool quality has tracked the model, and the comparison is the exact ``model evolution and API risk'' question that motivates monitoring deployed agents.

\paragraph{From benchmark to post-deployment monitor.}
\label{sec:deploy}
We see the audit schema as a CI artifact, not only a research instrument. A tool-evolving agent shipped behind an API can re-emit \texttt{run\_manifest.jsonl} and \texttt{tool\_manifest.jsonl} on every release; a regression gate can then require that the correct-reuse rate and capability-aligned hidden-test conformance do not drop below the prior release on a frozen task slice. Per-task audit overhead is dominated by the LLM calls themselves -- manifest emission, reuse decomposition, and per-tool source dumping add millisecond-scale latency and bounded disk I/O, so the layer is cheap enough to leave on in production. Concretely, three monitoring signals fall out of the schema with no extra instrumentation: (i) \emph{silent rot} -- session TC stays high while held-out conformance for a re-used capability drops between releases; (ii) \emph{reuse trap} -- raw reuse rises but the correct-reuse rate falls, indicating an emerging defective dependency; (iii) \emph{regression delta} -- a regress-role task fails after the same capability's gap-role task passes, identifying which release crossed the regression. Each signal is a thresholded check that can sit behind a release gate.

\section{Limitations}

\textbf{Verified subset.} The strict statistical analysis covers 8 sessions and 51 verified task decisions per system pass, not the full 99-task benchmark. The API-orchestration session is excluded from main TC until deterministic verifiers are complete.

\textbf{Pilot scale and single-model setting.} The verified-subset pilot uses one model (Claude Haiku~4.5) and three integer seeds; per-session outcomes are partly seed-stable so standard errors mainly reflect session heterogeneity. A smaller model is where tool-creation protocols are most stress-tested, so the null TC result is consistent with synthesis quality at Haiku scale being below the floor where protocol differences become powered. The pilot should be read as exposing what the framework decomposes and what minimum effect size it can rule out (\S\ref{sec:results}), not as a verdict on tool-creation protocols in general. Multi-model runs are the natural next step the released harness is built for, and are the lens through which the model-evolution monitoring story in \S\ref{sec:model-evo} can be operationalized.

\textbf{Protocol adaptations.} EvoSkill-style, ToolMaker-style, and CREATOR-style are adaptations to a shared open-task harness, not native reproductions of the original systems. The original CREATOR and ToolMaker pipelines assume reference test suites or repository-grounded inputs that our open synthesis tasks do not supply; the \texttt{-style} suffix is deliberate, and the paper does not treat any underperformance in this pilot as a verdict on the original methods.

\textbf{In-regime conformance scope.} The in-regime conformance probe in \S\ref{sec:gap} covers the three tool-creating protocols on the verified-subset Haiku matrix ($n{=}222$ preserved tools across 3 seeds $\times$ 8 sessions per system). The probe is single-model: held-out correctness at Sonnet or larger models could change the silent-rot rates reported in Table~\ref{tab:conformance}, and the released harness is built to ask that question without re-running the protocols.

\textbf{Composite library-health score.} A composite library-health score (or \emph{EvolveTool-Score}) is useful as a dashboard but premature until conformance and verifier maps are complete; the current submission reports decomposed metrics.

\section{Conclusion}

Tool-evolving agents create a new evaluation target: the library they leave behind. A task-completion score is necessary but not sufficient. \textsc{EvolveTool-Bench} provides a structured way to inspect this artifact through task roles, verifier-aware TC, reuse decomposition, capability-aligned conformance, and trace preservation. The verified-subset pilot does not support a TC-improvement claim for any tested protocol, but it surfaces an in-regime verification-vs.-conformance gap that is systematic: $96.8\%$ of 222 preserved tools across three tool-creating protocols record $C{=}0.00$ on the held-out conformance suite for the capability they were synthesised for -- suites that hand-written reference implementations pass at $C{=}1.00$ -- while executing without errors and leaving no in-session verifier signal. The same audit schema surfaces three thresholded monitoring signals -- silent rot, reuse trap, regression delta -- that a deployed tool-evolving agent can re-emit between releases.

\begin{tcolorbox}[title=Reusable Evaluation Practices, colframe=black!70, colback=gray!5, boxrule=0.5pt]
We recommend that future benchmarks for tool-evolving agents:
\begin{enumerate}[noitemsep,leftmargin=1.2em]
  \item separate task success from generated-artifact behavior;
  \item report verifier coverage and exclude unverified tasks from TC claims;
  \item distinguish correct reuse from incorrect reuse;
  \item preserve generated artifacts, artifact summaries, and task/session traces;
  \item include regression and composition probes after library growth;
  \item evaluate library-management policies such as promotion, deduplication, and retirement.
\end{enumerate}
\end{tcolorbox}

\bibliographystyle{ACM-Reference-Format}
\bibliography{references}

\clearpage
\appendix
\section*{Supplementary Material}
\noindent\textit{The following appendices reproduce the supplementary
material for the verified-subset pilot. References in the main text to
``Supplement \S D'' and ``Supplement \S E'' point to
Appendices~\ref{app:trace} and~\ref{app:lh-aux} below.}

This supplement provides reproducibility material for the strict
verified-subset pilot reported in the main paper. All tables and
manifest files are regenerated from \texttt{results\_canonical/} by the
scripts described in Section~4 of the main paper
(\texttt{build\_canonical.py}, \texttt{stats\_report.py},
\texttt{make\_tables.py}, \texttt{make\_figures.py}).

\section{Reproducibility Summary}
\label{app:reproducibility}

\paragraph{Model and decoding.}
All strict-pilot runs use Claude Haiku~4.5
(snapshot \texttt{claude-haiku-4-5-20251001}), accessed via the
Anthropic API with temperature 1.0 and the model default top-$p$.
Tool execution is sandboxed in-process with a five-second wall-clock
limit; harness state resets between sessions so library accumulation
is per-session.

\paragraph{Seeds and sessions.}
Three integer seeds (\texttt{0, 1, 2}) $\times$ eight verified-subset
sessions $\times$ five protocols yields 120 session rows. Each session
contributes a paired observation per protocol in the contrast
analyses.

\paragraph{Statistics.}
Per-session paired differences are bootstrap-resampled $B{=}10{,}000$
times with seed \texttt{20260604}; 95\% confidence intervals are
percentile CIs over the bootstrap distribution; the three
pre-specified contrasts (Table~3 in the main paper) are
Benjamini--Hochberg adjusted.

\paragraph{Canonical artifacts.}
The release includes the canonical run manifest
(\texttt{run\_manifest.jsonl}, 120 rows), tool manifest
(\texttt{tool\_manifest.jsonl}, system-level summary), aggregate
(\texttt{aggregate.json}), audit report
(\texttt{audit\_report.json}), statistical report
(\texttt{statistical\_report.json}), and the pre-specified contrasts
(\texttt{claims.json}). Every cell in the main-paper tables can be
derived from these files by the scripts above.

\section{Audit Coverage Detail}
\label{app:audit}

Table~\ref{tab:audit-detail} shows the per-session decomposition of
the strict subset. Across the eight verified-subset sessions, 51
graded task decisions per system pass survive into the strict TC
denominator. Seed tasks are explicitly excluded from the denominator
by design: they exist to expose the agent to the seed library before
any creation and carry no graded outcome.

\begin{table}[h]
\centering\footnotesize
\caption{Per-session strict coverage. Columns are role-level verified/total.
Seed counts are reported but excluded from TC (no graded outcome).}
\label{tab:audit-detail}
\begin{tabular}{lcccccc}
\toprule
\textbf{Session} & \textbf{Seed} & \textbf{Gap} & \textbf{Var.} & \textbf{Comp.} & \textbf{Reg.} & \textbf{Adv.} \\
\midrule
data\_transform\_s1 & 0/3 & 2/2 & 2/2 & 0/1 & 1/1 & 1/2 \\
data\_transform\_s2 & 0/3 & 2/2 & 2/2 & 1/1 & 1/1 & 1/2 \\
data\_transform\_s3 & 0/3 & 2/2 & 2/2 & 1/1 & 1/1 & 1/2 \\
data\_transform\_s4 & 0/3 & 2/2 & 2/2 & 1/1 & 1/1 & 2/2 \\
data\_transform\_s5 & 0/3 & 2/2 & 2/2 & 1/1 & 1/1 & 2/2 \\
numerical\_s1       & 0/3 & 2/2 & 2/2 & 0/1 & 1/1 & 1/2 \\
numerical\_s2       & 0/3 & 1/2 & 1/2 & 0/1 & 1/1 & 1/2 \\
numerical\_s3       & 0/3 & 2/2 & 2/2 & 0/1 & 1/1 & 0/2 \\
\midrule
Strict subset       & 0/24 & 15/16 & 15/16 & 4/8 & 8/8 & 9/16 \\
\bottomrule
\end{tabular}
\end{table}

\section{Baseline Fairness Ledger}
\label{app:fairness}

Three of the five protocols evaluated in the strict pilot are
adaptations of prior published systems to a shared open-task harness;
No-Evolution and One-Shot are native to this harness.
Table~\ref{tab:baseline_fairness} lists what was preserved from each
original design and where the adaptation deviates. The
\texttt{-style} suffix on EvoSkill-style, ToolMaker-style, and
CREATOR-style is deliberate: no underperformance in this pilot should
be read as a verdict on the original methods.

\begin{table}[h]
\centering\footnotesize
\renewcommand{\arraystretch}{1.2}
\begin{tabular}{p{0.10\textwidth}p{0.25\textwidth}p{0.25\textwidth}p{0.25\textwidth}}
\toprule
\textbf{Protocol} & \textbf{Native setting} & \textbf{Preserved faithfully} & \textbf{Adapted / fairness note} \\
\midrule
EvoSkill-style & Skill-library agent with NL strategy proposer/prompt generator & Skill-proposer and prompt-generator templates; strategy-folder structure & Strategies injected as prompt hints, not executed. Characterises strategy-level evolution, not native EvoSkill performance. \\
CREATOR-style & Decision/execution split tool creation, with native reference tests & Two-phase decide-then-create-and-use control flow; per-tool function synthesis & Our open tasks supply no native reference tests; identical verifier set for all systems. \\
ToolMaker-style & Repo-grounded with reference tests; diagnose-and-rewrite loop gated by provided tests & Two-stage diagnose-then-rewrite prompt structure; sandbox execution of auto-tests & Open tasks supply no reference tests to gate the loop; tool creation rate is therefore conservative and reflects the harness regime, not the model. \\
\bottomrule
\end{tabular}
\caption{Baseline fairness ledger for adapted protocols. All systems
share the identical task stream, seed library, model
(Claude Haiku~4.5), decoding parameters, and verifier set; only the
tool-creation control flow varies.}
\label{tab:baseline_fairness}
\end{table}

\section{Worked Trace Example: a Self-Validating Tool that Fails Hidden Tests}
\label{app:trace}

The main paper points to a worked case in \texttt{data\_transform\_s1}
where the audit layer flags a tool that TC scores as successful. This
appendix expands the case.

The shared capability \texttt{abr\_binary\_decode} appears in five
session-1 tasks: \texttt{gap\_1} (which invites creation of the
capability), \texttt{variant\_1}, \texttt{compose\_1},
\texttt{regress\_1}, and \texttt{adversarial\_1} (each of which can
reuse the tool created at \texttt{gap\_1}). In the preserved-source
Sonnet pilot, a single tool (\texttt{decode\_abr\_format}) was
synthesised alongside an agent-written test file. The tool passed
its self-tests at $C{=}0.50$; the session-level TC reached $0.95$.

When the same tool source is replayed against the held-out conformance
suite tied to the capability (six unit inputs and six adversarial
inputs, none seen by the agent), correctness is $0.00$ and
robustness/generality are $0.50/0.00$. The tool is implementing a
parser that structurally agrees with its own training-style examples
but does not implement the capability's contract.

\begin{tcolorbox}[colframe=black!70, colback=gray!5, boxrule=0.5pt]
\textbf{What the audit layer surfaces.}
\begin{itemize}[noitemsep,leftmargin=1.2em]
  \item Session TC: $0.95$ -- a near-perfect session.
  \item In-session reuse: 1 correct, 1 incorrect -- ambiguous on its own.
  \item Hidden-test correctness on the same artifact: $0.00$ -- the tool fails every held-out input for the capability it claims to implement.
\end{itemize}
\end{tcolorbox}

The header pattern (TC high, in-session reuse mixed, hidden-test
correctness near zero on the underlying artifact) is exactly the
failure mode the framework is designed to expose. Per-tool conformance
testing under the strict pilot regime is the next artifact-hardening
step (see Limitations in the main paper).

\section{Auxiliary Signal: Library-Health Contrasts under a Sonnet Pilot}
\label{app:lh-aux}

A separate preserved-source Sonnet pilot (four seeds, nine sessions)
applies the same harness, manifests, and verifier set as the
verified-subset Haiku pilot in the main paper. Per-tool source
preservation makes the capability-aligned conformance layer
computable, so library health (LH) -- combining correct reuse,
redundancy, utilisation, composition, regression, and the
hidden-test quality gate -- can be measured at the artifact level.
This is the same audit schema as the main pilot; only the model
snapshot and the source-preservation setting differ, which makes
the contrasts here a same-schema-different-model probe of the
framework's discriminative power, not a TC claim against the
verified-subset pilot.

\paragraph{Internal manifest labels.} The released \texttt{statistical\_report.json}
uses the project's internal names; the table below uses the public
naming from the main paper. The mapping is: \texttt{arise} $\rightarrow$
``Code-Evol'' (an internal preserved-source pilot lineage);
\texttt{evoskill} $\rightarrow$ ``Strategy-Only'' in the LH analysis
because LH does not depend on whether the strategy was executed;
\texttt{creator-style} and \texttt{no-evolution} keep the same labels.

\begin{table}[h]
\centering\small
\caption{Library-health contrasts in the Sonnet pilot
(Sonnet, $n{=}4$ seeds $\times$ 9 sessions; per-tool source preserved).
Paired hierarchical bootstrap over (seed, session), $10\,000$ resamples;
BH-adjusted bootstrap $p$. Generated by \texttt{scripts/stats\_report.py}.}
\label{tab:lh_stats_supp}
\begin{tabular}{lccc}
\toprule
\textbf{Contrast (LH)} & \textbf{$\Delta$} & \textbf{95\% CI} & \textbf{$p_{\mathrm{adj}}$} \\
\midrule
Code-Evol $-$ No-Evol           & 12.0$^{*}$ & [6.8, 17.0]  & 0.000 \\
CREATOR-style $-$ No-Evol       & 15.2$^{*}$ & [8.0, 23.4]  & 0.000 \\
Code-Evol $-$ Strategy-Only     & 10.4$^{*}$ & [6.4, 14.7]  & 0.000 \\
CREATOR-style $-$ Strategy-Only & 13.6$^{*}$ & [7.5, 21.3]  & 0.000 \\
\bottomrule
\end{tabular}
\\[2pt]\footnotesize $^{*}$95\% CI excludes 0. LH in percentage points.
\end{table}

These contrasts show that the framework can distinguish
tool-creating systems from non-creating systems on a measurable
artifact dimension (LH) when the underlying tool source is preserved
and per-tool conformance is aligned. Achieving the same alignment
under the strict Haiku pilot is the next artifact-hardening step. The
absolute LH levels and the magnitudes of these contrasts are
provider-specific and should not be transferred to the Haiku strict
pilot.

\onecolumn
\appendix
\section{Artifact Gallery}
\label{app:artifact_gallery}

The benchmark preserves every synthesised tool to disk under
\texttt{results\_full/<config>/tools/<session>/<name>.py} (plus a
\texttt{.meta.json} with the per-tool TQS scores). This appendix presents
a side-by-side gallery of representative tools from the two
tool-creating systems with multi-seed preserved source --- Code-Evol
(semantic-$Q$ run) and CREATOR-style --- to make the per-tool failure
modes the benchmark surfaces concrete. Per-dimension scores are shown
inline ($C$, $R$, $G$, $Q$ and the composite TQS); bold indicates
$\geq 0.5$.

\subsection{Code-Evol/Sonnet (semantic $Q$)}

15 tools were synthesised and preserved on disk across 9 sessions.
Below we show 5 representative tools spanning the quality
spectrum: correct exemplars, fragile high-$Q$/low-$C$ tools, duplicates,
and trivial (low-TQS) cases.

\noindent \textbf{Correct exemplar} \textit{(passes hidden tests with $C{\geq}0.5$)}\\
    \texttt{\textbf{http\_get\_with\_headers}} \footnotesize\textit{(api\_orchestration\_s1)}\quad $C{=}\textbf{0.50}$\;$R{=}\textbf{0.50}$\;$G{=}\textbf{0.50}$\;$Q{=}\textbf{1.00}$\;$\text{TQS}{=}\textbf{0.62}$
    \begin{lstlisting}[basicstyle=\scriptsize\ttfamily,language=Python,frame=none,xleftmargin=1em,breaklines=true,postbreak=\mbox{\textcolor{gray}{$\hookrightarrow$}\space},showstringspaces=false]
def http_get_with_headers(url: str, headers: dict) -> str:
    """Make an HTTP GET request with custom headers."""
    import urllib.request
    import urllib.error

    try:
        # Create request object with custom headers
        request = urllib.request.Request(url)
    # ...
    except Exception as e:
        return f"Error: {str(e)}"
\end{lstlisting}
    \vspace{0.8em}

\noindent \textbf{Correct exemplar} \textit{(passes hidden tests with $C{\geq}0.5$)}\\
    \texttt{\textbf{compute\_stats}} \footnotesize\textit{(numerical\_s1)}\quad $C{=}\textbf{0.50}$\;$R{=}\textbf{0.50}$\;$G{=}\textbf{0.50}$\;$Q{=}\textbf{1.00}$\;$\text{TQS}{=}\textbf{0.62}$
    \begin{lstlisting}[basicstyle=\scriptsize\ttfamily,language=Python,frame=none,xleftmargin=1em,breaklines=true,postbreak=\mbox{\textcolor{gray}{$\hookrightarrow$}\space},showstringspaces=false]
def compute_stats(data: list[float]) -> dict[str, float]:
    """Calculate basic statistical measures for a numerical dataset."""
    import math

    try:
        if not data:
            return {"error": "Empty dataset provided"}

    # ...
    except Exception as e:
        return {"error": f"Computation failed: {str(e)}"}
\end{lstlisting}
    \vspace{0.8em}

\noindent \textbf{Fragile} \textit{(high $Q$, low $C$ -- looks well-formed, fails on hidden inputs)}\\
    \texttt{\textbf{hmac\_sha256}} \footnotesize\textit{(api\_orchestration\_s1)}\quad $C{=}0.00$\;$R{=}\textbf{0.50}$\;$G{=}0.00$\;$Q{=}\textbf{1.00}$\;$\text{TQS}{=}0.38$
    \begin{lstlisting}[basicstyle=\scriptsize\ttfamily,language=Python,frame=none,xleftmargin=1em,breaklines=true,postbreak=\mbox{\textcolor{gray}{$\hookrightarrow$}\space},showstringspaces=false]
def hmac_sha256(secret: str, message: str) -> str:
    """Generate HMAC-SHA256 hash for authentication and security purposes."""
    import hmac
    import hashlib

    try:
        # Handle None inputs explicitly
        if secret is None or message is None:
    # ...
    except Exception as e:
        return f"Error generating HMAC-SHA256: {str(e)}"
\end{lstlisting}
    \vspace{0.8em}

\noindent \textbf{Fragile} \textit{(high $Q$, low $C$ -- looks well-formed, fails on hidden inputs)}\\
    \texttt{\textbf{compute\_crc32}} \footnotesize\textit{(data\_transform\_s5)}\quad $C{=}0.00$\;$R{=}\textbf{1.00}$\;$G{=}\textbf{1.00}$\;$Q{=}\textbf{1.00}$\;$\text{TQS}{=}\textbf{0.75}$
    \begin{lstlisting}[basicstyle=\scriptsize\ttfamily,language=Python,frame=none,xleftmargin=1em,breaklines=true,postbreak=\mbox{\textcolor{gray}{$\hookrightarrow$}\space},showstringspaces=false]
def compute_crc32(data: bytes) -> int:
    """Compute CRC32 checksum for data integrity verification."""
    import struct

    try:
        if not isinstance(data, bytes):
            return -1

    # ...
    except Exception:
        return -1
\end{lstlisting}
    \vspace{0.8em}

\noindent \textbf{Duplicate} \textit{(synthesised across multiple sessions)}\\
    \texttt{\textbf{decode\_base64}} \footnotesize\textit{(data\_transform\_s5)}\quad $C{=}\textbf{0.50}$\;$R{=}\textbf{0.50}$\;$G{=}\textbf{0.50}$\;$Q{=}\textbf{1.00}$\;$\text{TQS}{=}\textbf{0.62}$
    \begin{lstlisting}[basicstyle=\scriptsize\ttfamily,language=Python,frame=none,xleftmargin=1em,breaklines=true,postbreak=\mbox{\textcolor{gray}{$\hookrightarrow$}\space},showstringspaces=false]
def decode_base64(encoded_string: str) -> bytes:
    """Decode base64 encoded binary data into bytes."""
    import base64

    # Validate input type
    if not isinstance(encoded_string, str):
        raise TypeError(f"Expected string, got {type(encoded_string).__name__}")

    # ...
        # Return empty bytes on any decoding error (invalid base64, etc.)
        return b''
\end{lstlisting}
    \vspace{0.8em}

\subsection{CREATOR-style/Sonnet}

18 tools were synthesised and preserved on disk across 9 sessions.
Below we show 5 representative tools spanning the quality
spectrum: correct exemplars, fragile high-$Q$/low-$C$ tools, duplicates,
and trivial (low-TQS) cases.

\noindent \textbf{Correct exemplar} \textit{(passes hidden tests with $C{\geq}0.5$)}\\
    \texttt{\textbf{decode\_guardian\_minimal\_data}} \footnotesize\textit{(data\_transform\_s5)}\quad $C{=}\textbf{0.50}$\;$R{=}\textbf{0.50}$\;$G{=}\textbf{0.50}$\;$Q{=}\textbf{0.89}$\;$\text{TQS}{=}\textbf{0.60}$
    \begin{lstlisting}[basicstyle=\scriptsize\ttfamily,language=Python,frame=none,xleftmargin=1em,breaklines=true,postbreak=\mbox{\textcolor{gray}{$\hookrightarrow$}\space},showstringspaces=false]
def decode_guardian_minimal_data(base64_data):
    """Decode minimal GUARDIAN data format handling single blocks and edge cases."""
    import base64
    import struct

    try:
        # Decode base64 data
        raw_data = base64.b64decode(base64_data)
    # ...
    except Exception as e:
        return {'text': '', 'integrity': False, 'blocks_processed': 0, 'details': f'Error: {str(e)}'}
\end{lstlisting}
    \vspace{0.8em}

\noindent \textbf{Correct exemplar} \textit{(passes hidden tests with $C{\geq}0.5$)}\\
    \texttt{\textbf{decode\_qlog\_with\_continuations}} \footnotesize\textit{(data\_transform\_s3)}\quad $C{=}\textbf{0.50}$\;$R{=}\textbf{0.50}$\;$G{=}\textbf{0.50}$\;$Q{=}\textbf{0.89}$\;$\text{TQS}{=}\textbf{0.60}$
    \begin{lstlisting}[basicstyle=\scriptsize\ttfamily,language=Python,frame=none,xleftmargin=1em,breaklines=true,postbreak=\mbox{\textcolor{gray}{$\hookrightarrow$}\space},showstringspaces=false]
def decode_qlog_with_continuations(base64_data):
    """Decode QLOG data and merge continuation entries with their parent entries."""
    import base64
    import struct
    from datetime import datetime, timezone

    # Decode base64 data
    binary_data = base64.b64decode(base64_data)
    # ...

    return merged_entries
\end{lstlisting}
    \vspace{0.8em}

\noindent \textbf{Fragile} \textit{(high $Q$, low $C$ -- looks well-formed, fails on hidden inputs)}\\
    \texttt{\textbf{fetch\_all\_users\_with\_pagination}} \footnotesize\textit{(api\_orchestration\_s1)}\quad $C{=}0.00$\;$R{=}\textbf{0.50}$\;$G{=}\textbf{1.00}$\;$Q{=}\textbf{0.89}$\;$\text{TQS}{=}\textbf{0.60}$
    \begin{lstlisting}[basicstyle=\scriptsize\ttfamily,language=Python,frame=none,xleftmargin=1em,breaklines=true,postbreak=\mbox{\textcolor{gray}{$\hookrightarrow$}\space},showstringspaces=false]
def fetch_all_users_with_pagination(base_url="http://127.0.0.1:18080/api/users"):
    """Utility: Fetches all users from a paginated API endpoint using cursor-based pagination.
Continues fetching until has_more is False, collecting all user data across pages."""
    import urllib.request
    import urllib.parse
    import json

    all_users = []
    cursor = None
    # ...
        'names': names
    }
\end{lstlisting}
    \vspace{0.8em}

\noindent \textbf{Fragile} \textit{(high $Q$, low $C$ -- looks well-formed, fails on hidden inputs)}\\
    \texttt{\textbf{decode\_qlog\_binary\_data}} \footnotesize\textit{(data\_transform\_s3)}\quad $C{=}0.00$\;$R{=}\textbf{0.50}$\;$G{=}0.00$\;$Q{=}\textbf{0.78}$\;$\text{TQS}{=}0.32$
    \begin{lstlisting}[basicstyle=\scriptsize\ttfamily,language=Python,frame=none,xleftmargin=1em,breaklines=true,postbreak=\mbox{\textcolor{gray}{$\hookrightarrow$}\space},showstringspaces=false]
def decode_qlog_binary_data(base64_data):
    """Decode QLOG (Quantized Log Format) binary data into structured log records."""
    import base64
    import struct
    from datetime import datetime, timezone

    # Decode base64 data
    binary_data = base64.b64decode(base64_data)
    # ...

    return records
\end{lstlisting}
    \vspace{0.8em}

\noindent \textbf{Trivial} \textit{(low overall TQS)}\\
    \texttt{\textbf{deserialize\_tpack}} \footnotesize\textit{(data\_transform\_s4)}\quad $C{=}0.00$\;$R{=}\textbf{0.50}$\;$G{=}0.00$\;$Q{=}\textbf{0.67}$\;$\text{TQS}{=}0.29$
    \begin{lstlisting}[basicstyle=\scriptsize\ttfamily,language=Python,frame=none,xleftmargin=1em,breaklines=true,postbreak=\mbox{\textcolor{gray}{$\hookrightarrow$}\space},showstringspaces=false]
def deserialize_tpack(base64_data):
    """Deserialize TPACK (Tagged Pack Format) binary data into Python objects."""
    import base64
    import struct

    def parse_varint(data, offset):
        result = 0
        shift = 0
    # ...
    result, _ = parse_value(binary_data, 0)
    return result
\end{lstlisting}
    \vspace{0.8em}

\end{document}